\documentclass[11pt,a4paper]{article}

\usepackage[margin=2.5cm]{geometry}
\usepackage{lmodern}
\usepackage[T1]{fontenc}
\usepackage[utf8]{inputenc}
\usepackage{hyperref}
\usepackage{amsmath,amssymb}
\usepackage{microtype}
\usepackage{enumitem}

\title{Human versus Artificial Inteligence; \\
various significant examples in astrophysics}
\author{A. De R\'ujula${}^{a,b}$ \\
\\
${}^a$Instituto de F\'isica Te\'orica (UAM/CSIC), Univ. Aut\'onoma de Madrid, Spain;\\
${}^b$Theory Division, CERN, CH 1211 Geneva 23, Switzerland}

\newenvironment{qa}[2]{
  \subsection*{#1}
  \noindent\textbf{Question.} #2

  \medskip
  \noindent\textbf{Answer.}
}{\medskip}

\begin{document}
\maketitle





\begin{abstract}
In a recent arXiv posting [1] I reported the result of an experiment: asking Perplexity.ai to compare three items concerning (ordinary) Gamma Ray Burts (GRBs): the data, the standard paradigm(s) and the "Cannonball" (CB) model. Here I ask the same URL to extend this comparison to long--lasting GRBs, binary Neutron-Star mergers and their associated short--hard GRBs, low--luminosity GRBs, X--ray flashes, X--ray transients, and non--solar cosmic rays. The results of this experiment are enlightening but worrisome. Except for this abstract, two footnotes and two other references to standard [2] and CB-model [3] articles and talks, all of what follows is, verbatim, what the cited AI "opines".

\end{abstract}

\
\maketitle

\section{Introduction}
Discussion on GRB Models: Cannonball vs.$\!$ Fireball 

\section*{Overview}

This document collects a structured dialogue about gamma-ray burst (GRB)
models, focusing on the Cannonball (CB) model and the standard fireball
framework. The content is organized as a sequence of questions and
answers and includes a tabular comparison of model predictions versus
observations.

\section{Model Comparisons}

\begin{qa}{General comparison of GRB models}%
{Comparison between the cannonball and the standard model of GRBs.}
The cannonball (CB) model and the standard fireball/blast-wave model
offer different physical pictures of GRBs.
In the CB model, highly relativistic, compact plasmoids (``cannonballs'')
are ejected in narrow jets during core-collapse supernovae or similar
events, producing prompt gamma rays mainly by inverse Compton scattering
(ICS) of ambient photons, and afterglows via synchrotron radiation from
swept-up interstellar electrons.
In the standard fireball picture, a relativistically expanding outflow
drives internal and external shocks; prompt emission is typically
attributed to synchrotron or Compton processes in internal shocks or
photospheres, while afterglows arise from synchrotron radiation in an
external forward shock.
The CB model emphasizes analytic, few-parameter descriptions with strong
geometric and kinematic correlations (e.g.\ between peak energy and
luminosity), whereas the fireball framework is more flexible, with many
microphysical and geometric parameters tuned to reproduce the broad
diversity of light curves and spectra.
\end{qa}

\begin{qa}{Observable predictions distinguishing models}%
{Observable predictions distinguishing CB vs.\ fireball.}
In the prompt phase, the CB model predicts individual pulses that are
fast-rise, exponential-decay in time, with Band-like cutoff power-law
spectra whose peak energy and hardness track the pulse flux, as direct
consequences of ICS and relativistic beaming.
It also tends to predict high linear polarization for suitably aligned
sightlines, because scattering in a narrow jet geometry can produce
large polarization fractions.
The fireball framework reproduces Band-like spectra and a variety of
pulse shapes using internal shocks, magnetic dissipation, or
photospheric emission, but generally requires more detailed
microphysics and numerical modeling to obtain specific correlations and
polarization levels.
For afterglows, the CB model gives simple analytic light curves where
breaks and chromatic behavior follow from CB deceleration and viewing
angle evolution\footnote{Here Perplexity should have said ``Lorentz factor'',
not ``viewing angle''. Its only hallucination.}, while fireball afterglows rely on external-shock
synchrotron with possible energy injection, jet structure, and evolving
microphysics to explain canonical X-ray behavior, plateaus, jet breaks,
and chromatic deviations.
\end{qa}

\begin{qa}{Table: models vs.\ observations}%
{A table comparing the models with observations}
{\it See the table as the last page.}

\begin{table}[h]
\centering
\vspace{-1.2cm}
\small
\begin{tabular}{p{4cm}p{6cm}p{6cm}}
\hline
\textbf{Observable / test} &
\textbf{Cannonball (CB) model} &
\textbf{Fireball / ``standard'' model} \\
\hline

Prompt spectrum shape (Band-like, cutoff PL) &
Prompt ICS on ambient ``glory'' light gives an analytic cutoff power law
that reproduces Band-like spectra and their softening with time in many
bursts. &
Spectral shape is usually fit phenomenologically (Band, cutoff
power-law); synchrotron or thermal interpretations often require
microphysical tuning and can struggle with low-energy slopes in some
bright bursts. \\[0.4em]

Pulse shapes, $E_{\rm p}$--flux tracking &
Individual pulses from a single CB have predicted fast-rise,
exponential-decay shapes and hardness--intensity tracking, matching many
FRED-like GRB pulses. &
Internal-shock and photospheric implementations reproduce a wide variety
of pulse morphologies but with many free parameters; hardness--intensity
correlations are not a simple one-parameter consequence of geometry. \\[0.4em]

Prompt polarization (tens of \%) &
High polarization (tens of percent or more) is a generic expectation of
ICS in a narrow jet viewed off-axis, consistent with several claimed
high-polarization events (within substantial error bars). &
High polarization is possible but typically requires ordered magnetic
fields or special viewing geometries; early reports of very high
polarization were challenging for simple internal-shock scenarios. \\[0.4em]

Early X-ray ``steep--flat--steep'' canonical afterglow &
Interpreted as a transition from ICS-dominated tails to synchrotron
emission from a decelerating CB; simple analytic light curves can fit
many Swift X-ray afterglows with few parameters. &
Initially unexpected; now modeled with combinations of energy injection,
evolving microphysics, high-latitude emission and jet structure,
introducing several additional ingredients beyond the basic external
shock. \\[0.4em]

Jet breaks in X-ray/optical afterglows &
Breaks arise from CB deceleration and viewing-angle evolution; predicted
correlations between break time and other afterglow properties are
reported to agree with several samples. &
Classic top-hat jet models predicted achromatic jet breaks; Swift showed
many missing or chromatic breaks, pushing toward structured jets and
energy injection, and weakening the original clean jet-break test. \\[0.4em]

GRB--SN association (long GRBs) &
Long GRBs are expected to be accompanied by SN-like bumps; the CB model
anticipated GRB--SN links and fits multiple GRB/SN light curves with a
single SN template plus CB afterglow. &
Long GRBs are associated with core-collapse SNe in collapsar-type
fireball models; many SN bumps are reproduced with separate SN
templates and flexible energy injection histories in the afterglow. \\[0.4em]

Amati and related correlations &
Amati-type $E_{\rm p}$--$E_{\rm iso}$ relations arise almost directly
from jet Lorentz and Doppler factors; slopes and scatter follow from
kinematics and viewing-angle distributions. &
Amati and similar correlations are reproduced empirically and are often
treated as population trends; several physical explanations (e.g.\
photospheric emission, internal shocks) coexist without a single minimal
derivation. \\[0.4em]

Prompt optical flashes and early multi-band SEDs &
Early optical emission is attributed to ICS or early synchrotron from
the same CBs; fits to bursts with prompt optical data show consistent
spectral energy distributions and temporal evolution across bands. &
External-shock synchrotron explains many early optical light curves, but
very bright optical flashes or chromatic plateaus often require extra
components such as reverse shocks or late energy injection. \\[0.4em]

Late radio afterglow and apparent superluminal motion &
Predicts apparent superluminal motion of the radio image for nearby
events; modeled centroid shifts and radio light curves can agree within
errors for specific bursts. &
Jet geometry can also yield apparent superluminal motion, but standard
afterglow work usually emphasizes light curves rather than detailed
centroid motion; explicit fits of motion exist only for a few cases. \\[0.4em]

Overall parameter economy vs.\ data volume &
Emphasizes a small set of geometrical and environmental parameters to
fit prompt emission, afterglow, and associated SNe, stressing analytic,
falsifiable relations. &
Provides a very flexible framework (internal and external shocks,
magnetic and thermal components, structured jets); capable of fitting
diverse data but with many adjustable microphysical and geometric
parameters. \\
\hline
\end{tabular}
\caption{Qualitative comparison of the CB and fireball models against
representative observational features of GRBs.}
\end{table}

\end{qa}

\section{Sociology and model flexibility}

\begin{qa}{Why proponents continue to work on the standard model}%
{Why do defenders of the ``standard'' model continue to work on it?}
Proponents argue that, despite its shortcomings, the fireball framework
has substantial empirical success: external-shock synchrotron afterglows
explain many broadband light curves, and numerous ``textbook'' GRBs were
first interpreted within this picture.
The model is adaptable; when Swift and later missions revealed plateaus,
flares, chromatic breaks and extended central-engine activity, the
community extended the framework (structured jets, energy injection,
evolving microphysics) rather than abandoning it.
Because the formalism ties directly into relativistic shock physics,
MHD, and simulations of collapsars and mergers, it integrates well with
other areas of high-energy astrophysics, which reinforces its status as
the default language for data analysis.
Pedagogically and sociologically, the fireball picture dominates reviews,
schools, and grant culture, so incremental refinements within this
paradigm are institutionally safer and more familiar than shifting to a
minority alternative such as the CB model.
\end{qa}

\begin{qa}{Justifying ongoing parametrization tweaks}%
{How do proponents justify continued parametrization tweaks in the model?}
Advocates typically present parameter evolution and added degrees of
freedom as data-driven refinement: as higher-quality and broader-band
observations appear (including GeV and TeV detections), previously
adequate fixed microphysical parameters are seen as oversimplified,
motivating time- or environment-dependent $\epsilon_e$, $\epsilon_B$,
or electron index $p$.
They argue that collisionless shocks and turbulent magnetic fields are
intrinsically complex, so a truly realistic model must allow effective
parameters to vary rather than enforcing universal constants.
Global fits to large GRB samples are then used to claim that, even with
extra flexibility, parameter distributions remain clustered in plausible
ranges, so the framework is not completely unconstrained.
Conceptually, deviations from simple closure relations or spectra are
reinterpreted as diagnostics of circumburst environment and microphysics
rather than as falsifications of the core external-shock paradigm,
which legitimizes introducing additional parametrization while still
calling the model ``standard''.
\end{qa}

\section{Beyond GRBs}

This section summarizes how various high--energy transient classes and
non--solar cosmic rays are understood in the standard fireball model and
in the Cannonball (CB) model.  The classes considered are: classical
long GRBs, ultra--long / long--lasting GRBs, short--hard GRBs (SHBs),
low--luminosity GRBs (LL GRBs), X--ray flashes (XRFs), X--ray
transients (XRTs), and Galactic and extragalactic non--solar cosmic
rays.

For each class, we briefly list the key observational properties and the
interpretation in both theoretical frameworks.


\section{Long Gamma--Ray Bursts (classical, high--luminosity)}

\subsection*{Observational properties}


\begin{itemize}[noitemsep]
  \item Duration\footnote{Traditionally events with $\gtrsim 2$ s duration are simply called GRBs. Here  Perplexity.ai discusses them again.  I left its responses untouched, avoiding to "talk" to it and only asking questions, as stated in the Abstract.}
  typically $\gtrsim 2$ s; prompt emission with Band--like
        spectra and peak energies $\sim 10^2$--$10^3$ keV.
  \item Strong association with broad--line Type Ic supernovae at low
        redshift.
  \item Empirical correlations such as Amati/Yonetoku relations between
        spectral peak energy and energetics.
  \item X--ray afterglows often showing a canonical ``steep--flat--steep''
        behaviour; optical and radio afterglows with jet breaks in some
        events.
\end{itemize}

\subsection*{Standard fireball interpretation}

Long GRBs are produced by relativistic jets launched in collapsar
events.  Prompt emission is attributed to internal shocks, magnetic
dissipation and/or photospheric processes in the jet; afterglows are
synchrotron radiation from an external forward shock in the surrounding
medium.  The framework explains many broadband afterglows but often
requires evolving microphysics, structured jets and energy injection to
account for plateaus, chromatic breaks and detailed spectral evolution.

\subsection*{Cannonball interpretation}

In the CB model, long GRBs arise from highly relativistic
plasma clumps (``cannonballs'') ejected in ordinary core--collapse
supernovae.  Prompt $\gamma$--rays are produced mainly by inverse
Compton scattering (ICS) of ambient photons by electrons inside the
CBs, and afterglows are synchrotron radiation from swept--up
interstellar electrons.  Band--like spectra, pulse shapes and
Amati--type correlations follow from the kinematics of CBs and their
Doppler factors, and many Swift X--ray/optical light curves are fit
with simple analytic forms using few parameters.

\section{Ultra--Long / Long--Lasting GRBs}

\subsection*{Observational properties}

\begin{itemize}[noitemsep]
  \item Prompt emission lasting $\sim 10^3$--$10^4$ s or more, sometimes
        with long soft tails.
  \item Otherwise similar to long GRBs in spectra, afterglows and, in
        some cases, supernova associations.
\end{itemize}

\subsection*{Standard fireball interpretation}

Ultra--long GRBs are treated as a rare subclass of long GRBs with an
unusually long--lived central engine.  Proposed mechanisms include
fallback accretion on a black hole, magnetar spin--down or tidal
disruption--like scenarios.  In some variants a very low ambient density
delays jet deceleration.  Extended energy injection is used to explain
long prompt phases and X--ray plateaus, but the basic fireball
afterglow picture is retained.

\subsection*{Cannonball interpretation}

In the CB model, long--lasting prompt emission is attributed to
prolonged ejection of CBs or to an extended accretion phase.  Once CBs
are launched, their ICS and subsequent synchrotron emission follow the
same analytic laws as for ordinary long GRBs.  Very diverse Swift
broad--band light curves, including long--lasting and plateau--rich
events, are reproduced by varying a small number of CB and environmental
parameters, without invoking a fundamentally new GRB class.

\section{Short--Hard GRBs (SHBs)}

\subsection*{Observational properties}

\begin{itemize}[noitemsep]
  \item Prompt duration typically $\lesssim 2$ s with hard spectra.
  \item Some events are firmly associated with binary neutron--star
        mergers (e.g.\ GW170817/GRB 170817A).
  \item Afterglows consistent with structured jets seen on-- or
        off--axis; optical/IR kilonova/macronova counterparts in several
        cases.
\end{itemize}

\subsection*{Standard fireball interpretation}

SHBs arise from compact object mergers launching relativistic jets.
The prompt emission is produced by internal dissipation in the jet, and
the afterglow is a standard external--shock fireball.  The kilonova is a
separate component powered by radioactive decay of r--process ejecta.
The structured--jet afterglow of GRB~170817A is taken as strong support
for this picture.

\subsection*{Cannonball interpretation}

In the CB framework, SHBs and some SN--less GRBs originate in systems
containing neutron stars.  CBs launched in such systems produce the
prompt $\gamma$--rays and subsequent afterglows via the same ICS and
synchrotron mechanisms as long GRBs.  The optical and X--ray afterglows
are attributed to neutron--star--powered CB activity, with the aim of
describing SHBs and events like GRB~170817A using the same basic CB
machinery.

\section{Low--Luminosity GRBs (nearby low--L bursts)}

\subsection*{Observational properties}

\begin{itemize}[noitemsep]
  \item Nearby ($z \lesssim 0.1$), much lower isotropic luminosities than
        classical long GRBs.
  \item Strong association with Type Ic supernovae.
  \item High local rate compared to that inferred for high--luminosity
        long GRBs.
\end{itemize}

\subsection*{Standard fireball interpretation}

Many studies treat LL GRBs as a distinct population, for example
shock--breakout events or mildly relativistic jets with wide opening
angles and different energetics.  Their high local rate and inferred
beaming factors often favour a separate class rather than simply
off--axis views of high--luminosity GRBs.

\subsection*{Cannonball interpretation}

In the CB model, LL GRBs are ordinary CB GRBs seen far off--axis.  The
same CB kinematics, Doppler factors and ICS mechanism are used as for
classical long GRBs, with the lower apparent luminosities and softer
spectra arising primarily from viewing angle.  Thus LL GRBs are unified
with high--luminosity GRBs through geometry rather than through a
distinct physical engine.

\section{X--ray Flashes (XRFs)}

\subsection*{Observational properties}

\begin{itemize}[noitemsep]
  \item Softer spectra and lower peak energies than typical long GRBs.
  \item Often lower apparent luminosity, but with hosts and (when seen)
        supernovae similar to long GRBs.
\end{itemize}

\subsection*{Standard fireball interpretation}

XRFs are interpreted in a variety of ways: some are modeled as off--axis
GRBs or low--Lorentz--factor jets, while others are considered a
somewhat different subclass requiring modified jet structure or
photospheric conditions.  There is no single, universally adopted
fireball explanation, and population studies often leave room for both
off--axis and intrinsically soft scenarios.

\subsection*{Cannonball interpretation}

Within the CB model, XRFs are by construction far off--axis long GRBs.
The same CB kinematics and ICS mechanism apply; increasing viewing angle
shifts the observed peak energy to lower values and reduces the apparent
luminosity.  XRFs therefore belong to the same underlying CB population
as classical long GRBs.

\section{X--ray Transients (XRTs / fast extragalactic X--ray transients)}

\subsection*{Observational properties}

\begin{itemize}[noitemsep]
  \item Soft, sometimes short--lived extragalactic X--ray events with a
        range of luminosities and durations.
  \item A subset shows properties consistent with off--axis GRB
        afterglows, but the overall sample is heterogeneous and not yet
        fully classified.
\end{itemize}

\subsection*{Standard fireball interpretation}

In the fireball framework, some XRTs are modeled as afterglows of
off--axis GRBs from structured jets.  Detailed simulations indicate that
only a limited region of parameter space reproduces the observed
luminosity--duration--slope distributions, so a pure off--axis GRB
origin is plausible for part of the population but not for all events.
Other XRTs may require different central engines or environments.

\subsection*{Cannonball interpretation}

CB analyses identify a subset of XRTs as neutron--star--powered
afterglows produced by CBs, extending the same CB/NS machinery used for
short GRBs and SN--less GRBs.  The goal is again a unified CB-based
origin for GRB--like X--ray transients, with diversity arising from
engine type, viewing angle and environment rather than from unrelated
transient families.

\section{Non--Solar Cosmic Rays}

\subsection*{Observational properties}

\begin{itemize}[noitemsep]
  \item A nearly featureless power--law spectrum over many decades in
        energy, with breaks such as the knee and ankle.
  \item Composition evolving from light to heavier nuclei with energy.
  \item Small but nonzero anisotropies and correlations with
        star--forming regions and non--thermal emission in galaxies.
\end{itemize}

\subsection*{Standard paradigm interpretation}

In the generally accepted picture, Galactic cosmic rays are accelerated
mainly by diffusive shock acceleration in supernova remnants, with
additional contributions from pulsars, superbubbles and extragalactic
sources at the highest energies.  After acceleration, particles undergo
diffusive propagation in the Galactic magnetic field, with
energy--dependent escape and interactions with interstellar gas and
radiation.  This source-plus-propagation framework is implemented in
standard cosmic--ray transport codes and underlies fits to the observed
spectra, composition, anisotropy and diffuse $\gamma$--ray emission of
galaxies.

GRBs may contribute to ultra--high--energy cosmic rays in some models,
but ordinary Galactic cosmic rays are not tightly linked to GRB
phenomenology in this paradigm.

\subsection*{Cannonball interpretation}

The CB model proposes that the same CBs which produce GRBs also generate
essentially all non--solar cosmic rays across the full observed energy
range.  As CBs plow through ambient interstellar matter and radiation,
they elastically scatter and Fermi--accelerate particles; with suitable
CB parameters and Galactic conditions, the resulting spectra and
abundances are claimed to reproduce the observed power laws, knee/ankle
features and composition with relatively few free parameters.  In this
view, GRBs, long--lived CBs and cosmic rays are different manifestations
of a single underlying process, rather than loosely related
phenomena.

\section{Synthesis}

Across long GRBs, ultra--long GRBs, low--luminosity GRBs, XRFs, many
XRTs and non--solar cosmic rays, the CB model promotes a single unified
jet picture: cannonball jets from supernovae and neutron--star systems,
with observed classes distinguished primarily by engine duration,
viewing angle, ambient density and propagation effects.  The standard
fireball and cosmic--ray paradigms successfully explain many afterglows,
the merger--kilonova--short--GRB connection, and a wide range of
cosmic--ray data, but tend to treat ultra--long GRBs, LL GRBs, some
XRFs, parts of the XRT population and cosmic rays as separate or loosely
connected subclasses, each requiring specific sources, jet structures or
propagation conditions within a broader, more flexible theoretical
framework.

{}

\end{document}